\documentclass[11pt]{article}
\usepackage{PRIMEarxiv}
\usepackage{geometry}
\usepackage{amsmath,amssymb}
\usepackage{graphicx}
\usepackage{multirow}
\usepackage{booktabs}
\usepackage{hyperref}
\usepackage{subcaption}
\usepackage{authblk}
\usepackage{array}
\usepackage{listings}
\usepackage[dvipsnames]{xcolor}

\usepackage{fancyhdr}
\usepackage{adjustbox}

\fancypagestyle{firstpage}{
  \fancyhead{}
  \lhead{
      \adjustbox{valign=m}{\includegraphics[height=0.9cm]{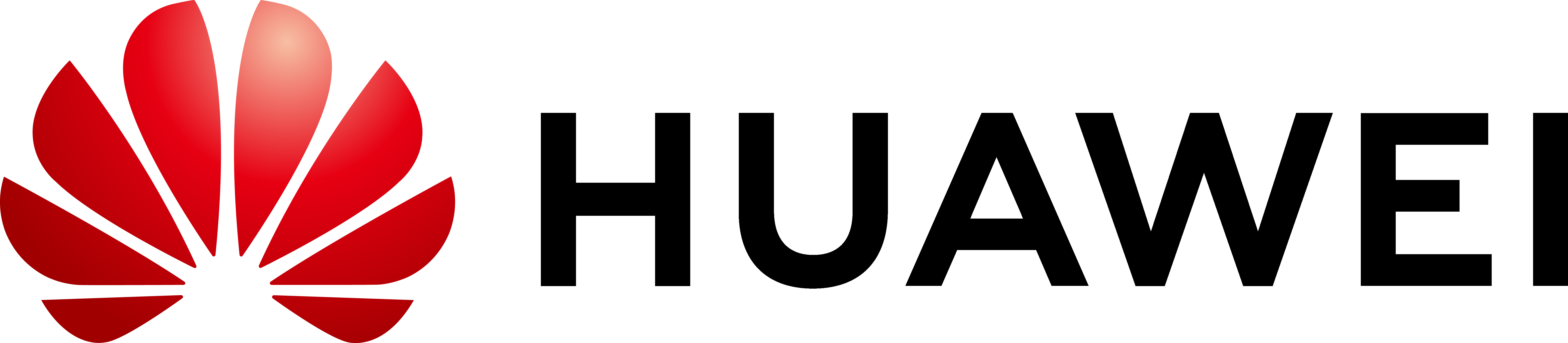}}
      \hspace{1em}
      \adjustbox{valign=m}{\includegraphics[height=0.86cm]{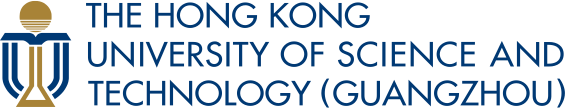}}
      \hspace{1em}
      \adjustbox{valign=m}{\includegraphics[height=0.9cm]{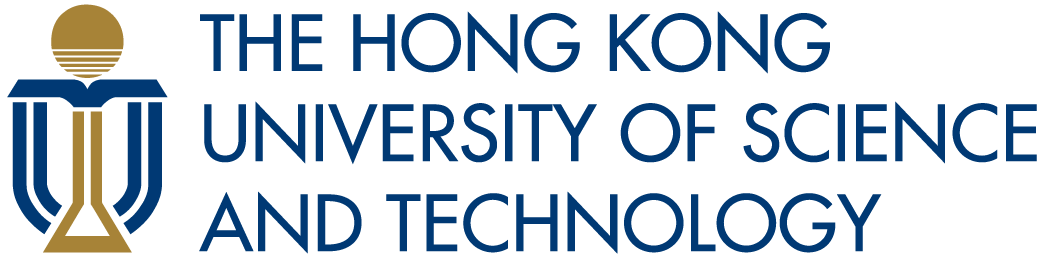}}
  }
  
  \setlength{\headsep}{1.3cm}
}

\title{HyperParallel: A Supernode-Affinity AI Framework\footnote{The full list of contributors is presented in the Acknowledgment Section at the end of the paper.}}
\author[$\dagger$]{Xin Zhang}
\author[$\ddagger$]{Beilei Sun}
\author[$\ddagger$]{Teng Su}
\author[$\ddagger$]{Qinghua Zhang}
\author[$\ddagger$]{Chong Bao}
\author[$\dagger$]{\\Lei Chen}
\author[$\ddagger$]{Xuefeng Jin}

\affil[$\dagger$]{\textit{HKUST and HKUST(GZ), Hong Kong and Guangzhou, China}}
\affil[$\ddagger$]{\textit{Huawei Technologies Co., Ltd., China}}

\date{}

\begin{document}
\maketitle

\thispagestyle{firstpage}

\begingroup
\renewcommand\thefootnote{*}
\footnotetext{The full list of contributors is presented in the Acknowledgments Section at the end of the paper.}
\endgroup

\begin{abstract}
The emergence of large-scale, sparse, multimodal, and agentic AI models has coincided with a shift in hardware toward supernode architectures that integrate hundreds to thousands of accelerators with ultra-low-latency interconnects and unified memory pools. However, existing AI frameworks are not designed to exploit these architectures efficiently, leading to high programming complexity, load imbalance, and poor memory utilization. In this paper, we propose a supernode-affinity AI framework that treats the supernode as a single logical computer and embeds hardware-aware orchestration into the framework. Implemented in MindSpore, our HyperParallel architecture comprises HyperOffload for automated hierarchical memory management, HyperMPMD for fine-grained MPMD parallelism across heterogeneous workloads, and HyperShard for declarative parallel strategy specification. Together, these techniques significantly improve training and inference efficiency while reducing parallel programming and system tuning overhead, demonstrating the necessity of supernode affinity for next-generation AI frameworks.
\end{abstract}

\section{Introduction}
The rapid advancement of artificial intelligence has been driven by the co-evolution of model architectures, training paradigms, and hardware systems~\cite{mudigere2022software,pandey2019machine, paszke2019pytorch}. Modern foundation models continue to scale in both parameter count~\cite{fedus2022switch,hoffmann2022training,ren2023pangu} and functional scope, expanding from language-only systems to multimodal~\cite{sun2024generative}, sparse~\cite{chen2022towards,dettmers2023spqr}, and agentic architecture~\cite{yu2025dapo}. These models introduce significant heterogeneity in computational loads, irregular data routing, and massive demands for memory and communication bandwidth. Simultaneously, hardware architectures have evolved into supernode-integrated clusters featuring thousands of accelerators connected via ultra-low-latency, peer-to-peer interconnects and unified memory pools~\cite{li2023pond,ubmesh,matrix384}. In this landscape, the AI framework has emerged as a critical middleware layer, bridging high-level algorithmic abstractions with increasingly complex hardware architectures.

We find that existing AI frameworks face a \textit{performance-programmability gap}. Traditional Single Program, Multiple Data (SPMD) paradigms~\cite{darema2001spmd} struggle to handle the load imbalances inherent in Mixture-of-Experts (MoE)~\cite{chen2022towards,dettmers2023spqr} and multimodal architectures. Furthermore, the manual management of intermediate states (weights, activations, and KV caches) across hierarchical memory layers imposes a prohibitive engineering burden on researchers. Current system extension layers~\cite{barham2022pathways,jax2018github,jiang2024megascale,jung2025hlx,rasley2020deepspeed,ren2021zero,shoeybi2019megatron}, while effective in specific niches, often lead to tight coupling between model logic and hardware topology, resulting in poor code portability and suboptimal cluster utilization.

These challenges suggest that incremental optimizations to existing frameworks are insufficient. Instead, AI frameworks must undergo an architectural shift that explicitly embraces the characteristics of supernodes and rethinks how parallelism, memory, and execution are abstracted. In this paper, we argue that supernode affinity should be a core design principle of next-generation AI frameworks. A supernode-affinity framework treats the supernode as a single, logically unified computing entity and embeds hardware-aware orchestration directly into the framework runtime and compiler. By doing so, it can shield algorithm developers from low-level system complexity while systematically unlocking the performance potential of modern hardware.

Based on this insight, we propose a supernode-affinity AI framework and present its design and implementation within MindSpore. The framework introduces a novel HyperParallel architecture that integrates three key components: HyperOffload, which decouples computation from model state through unified memory pooling and automated offloading; HyperMPMD, which extends parallel execution from SPMD to fine-grained Multiple Program, Multiple Data (MPMD) to address heterogeneous and multi-task workloads; and HyperShard, which provides a declarative parallel programming interface that decouples algorithm logic from parallel strategy design. Together, these components enable flexible, high-efficiency execution for MoE, omni-modal, and reinforcement learning workloads on supernode clusters.

This paper makes the following contributions:
\begin{itemize}
    \item Systematic Analysis of Emerging Challenges: We analyze recent trends in model workloads, hardware architectures, and training–inference paradigms, and identify the fundamental limitations they impose on existing AI frameworks when deployed on supernodes.
    \item Supernode-Affinity Framework Design: We propose a new architectural paradigm that treats the supernode as a giant computer, abstracting complex network topologies, hierarchical memory systems, and heterogeneous execution environments within the framework.
    \item HyperParallel Architecture: We introduce and implement HyperOffload, HyperMPMD, and HyperShard, enabling unified state management, flexible MPMD parallelism, and declarative parallel programming.
    \item Empirical Validation: We demonstrate that the proposed framework significantly improves training and inference efficiency while reducing development and optimization overhead, validating its effectiveness on representative large-scale workloads.
\end{itemize}

The remainder of this paper is organized as follows. Section~\ref{sec2} characterize the evolving of models and hardware, reviews existing AI frameworks, and analyzes the challenges faced by AI frameworks in the trillion-scale LLMs era. Section~\ref{sec4} presents the design and key technologies of the proposed supernode-affinity framework. Section~\ref{sec5} concludes with a discussion of results and future directions.

\section{Background}
\label{sec2}

The AI framework serves as a critical middleware layer situated between model algorithms and underlying hardware architectures, acting as the foundational software infrastructure that accelerates progress across the artificial intelligence domain. Its evolution is intrinsically driven by the dual requirements of algorithmic complexity and hardware innovation. On one hand, the framework must provide a high-level abstraction that enables researchers to efficiently develop, debug, and validate novel architectures, thereby fostering rapid algorithmic iteration. On the other hand, it must systematically transform these high-level abstractions into optimized executable programs that maximize hardware utilization and unlock the full potential of specialized computational resources.

\subsection{Small Deep Learning Models Era}

\textbf{Models and Hardware.} In the early stages of deep learning, most models are small in size. For example, small-scale Computer Vision (CV) models~\cite{zhao2024review} usually fit within a single accelerator's memory. Natural Language Processing (NLP) models~\cite{liu2019roberta} with modest parameter counts, while exceeding single-card capacity, usually fit within a single machine with multiple GPUs.

\noindent \textbf{Training Paradigms.} When training small deep learning models, the dominant training strategy for CV models is Data Parallelism (DP), with single-card computational power serving as the primary performance bottleneck. For modest-sized NLP models, model parallelism is usually employed to shard components across multiple cards within a node. High intra-node communication volume makes internal bandwidth the critical bottleneck.

\noindent \textbf{AI Frameworks.} To handle such workloads on a single machine, PyTorch~\cite{paszke2019pytorch} emerged as a dominant ecosystem with high execution flexibility. PyTorch's core strengths lie in its dynamic computational graph and robust ecosystem, making it ideal for rapid prototyping, model customization, and academic research.

\subsection{Billion-scale Large Language Models Era}

\noindent \textbf{Models and Hardware.} With the appearance of Large Language Models, the model parameters quickly expand to hundreds of billions. The massive volume of the pretraining dataset also poses higher requirements on the hardware. Typically, server clusters containing multiple nodes are built when training billion-scale Large Language Models.

\noindent \textbf{Training Paradigms.} Pretraining and supervised finetuning are the two most common training tasks for billion-scale LLMs. And multi-dimensional sharding (Data, Tensor, and Pipeline Parallelism) strategies are widely adopted. Different sharding strategies impose varied demands on cluster bandwidth. In particular, Tensor Parallelism (TP), Pipeline Parallelism (PP), and Context Parallelism (CP) necessitate frequent cross-server communication—often exceeding several GBs per transfer—which is difficult to hide via computation-communication overlapping. For example, the data traffic overhead of TP accounts for 52.9\% training time in a typical training setting~\cite{ubmesh}. Consequently, cross-machine communication becomes the primary performance bottleneck, leading to suboptimal cluster resource utilization.

\noindent \textbf{AI Frameworks.} PyTorch's primary limitations include the lack of comprehensive Multiple Program, Multiple Data (MPMD) capabilities and the absence of native support for the disaggregation of computation and storage. To address the bottlenecks, many frameworks, such as JAX~\cite{jax2018github}, DeepSpeed~\cite{rasley2020deepspeed}, Megatron~\cite{shoeybi2019megatron}, and HLX~\cite{jung2025hlx}, are developed as specialized system extension layers. These frameworks focus on distributed strategies, memory optimization, parallel topologies, hybrid attention compatibility, and high-efficiency multimodal processing.

\begin{itemize}
    \item \textbf{JAX: Functional Programming and Compiler-Driven Acceleration} -- JAX utilizes a functional programming paradigm combined with composable transformations (\textit{jit, pmap, vmap}), enabling the compiler to perform unified optimization across operators, parallelism, and differentiation. It excels in large-scale TPU training, scientific computing, and research requiring precise control over computational graph transformations. However, its ecosystem is highly specialized, leading to elevated programming complexity.
    
    \item \textbf{DeepSpeed: A \textit{System Toolbox} for Large-Scale Training} -- DeepSpeed provides the ZeRO series of redundancy-elimination sharding, MoE optimizers, efficient communication, and offload support, serving as critical infrastructure for training ultra-large models under constrained GPU resources. It acts as a system enhancement layer tightly coupled with PyTorch to compensate for deficiencies in memory efficiency and distributed scheduling. A significant drawback is its assumption of relatively uniform network topologies; in actual supernodes where inter-rack latency and bandwidth vary drastically, DeepSpeed's reliance on SPMD and static sharding can lead to severe load imbalances and hardware underutilization.
    
    \item \textbf{Megatron: Extreme Parallelism for Dense and Sparse Models} -- Megatron emphasizes the optimization of pipeline parallelism, tensor parallelism, and MoE parallelism to deliver peak distributed performance. Its "scale-first" design philosophy is ideal for environments with abundant GPU resources pursuing maximum throughput. While Megatron-Core is evolving into a foundational module for LLM training, the framework suffers from tight coupling between communication logic and hardware topology. This results in poor code portability, high programming complexity, and prohibitive costs for parameter tuning and performance optimization.

    \item \textbf{HLX: Acceleration for Hybrid Attention Architectures} -- HLX is designed for novel Transformer variants featuring linear and global hybrid attention. It introduces multi-threaded parallel orchestration and dual-dimensional (row/column) computational pipelining to accelerate sparse and linear attention operators. By deeply integrating operator structure with thread scheduling, HLX provides a system-level acceleration solution for hybrid attention models. However, it lacks framework-level unified memory pool management for supernodes, forcing researchers to manually manage intermediate training and inference states.
\end{itemize}

\noindent \textbf{Related Research Works.} Besides the frameworks discussed above, there are also some research works addressing other bottlenecks within the entire life-cycle of LLM pretraining.

\begin{itemize}
    \item \textbf{Cluster-level Fault Tolerance and Algorithm-System Co-design} -- {MegaScale}~\cite{jiang2024megascale} represents the trend in cluster-level training infrastructure. In environments with massive distributed computing power spanning multiple nodes, it achieves high Model Flops Utilization (MFU) on 10,000-card clusters through algorithm/kernel optimization, 3D parallel communication overlapping, and automated fault monitoring. Nevertheless, achieving effective computation-communication overlap requires highly customized parallel strategies and code implementations, leading to prohibitive debugging and migration costs.
    
    \item \textbf{Sharding and Offloading for Mid-to-Large Scale Training} -- Exemplified by the \textit{ZeRO} ecosystem (specifically ZeRO-3 and ZeRO-Offload)~\cite{rajbhandari2020zero, ren2021zero}, this direction targets the training of models ranging from tens of billions to trillions of parameters. The core approach involves the full sharding of parameters, gradients, and optimizer states across GPU clusters, often combined with CPU/NVMe offloading to break GPU memory walls. These methods utilize efficient communication partitioning to achieve near-linear scaling. However, they rely on static memory partitioning and lack automated management capabilities for unified memory pools, leading to potential memory fragmentation.

    \item \textbf{Scheduling and Universal Execution Layers for Ultra-Large Accelerator Clusters} -- \textit{Pathways}~\cite{barham2022pathways} demonstrates the necessity of a unified execution layer for ultra-large clusters. This direction focuses on asynchronous dataflow, scheduling, and topology optimization across accelerator pods to enable seamless collaboration between pipeline, data, and model parallelism. Unlike traditional rigid SPMD models, it supports fine-grained, irregular routing essential for MoE models. However, its reliance on a centralized resource manager and coordinator can introduce scheduling overhead and latency bottlenecks when managing highly dynamic, massive-scale task graphs across heterogeneous nodes.
    
\end{itemize}

\subsection{Trillion-Scale Large Language Models Era}

\noindent \textbf{Models.} Recently, we have observed major changes in the scale, architecture, and target modality for mainstream large language models. Concretely, the state-of-the-art LLMs usually have trillion-scale parameters and utilize Mixture-of-Experts (MoE) architectures to increase sparsity when scaling parameters. The support for multiple modalities beyond the text also prevails.

\begin{itemize}
    \item From the perspective of model sizes and architectures, many state-of-the-art LLMs, such as DeepSeek-V3~\cite{liu2024deepseek} and Qwen-3~\cite{yang2025qwen3}, utilize Mixture-of-Experts (MoE) architectures to increase sparsity when scaling parameters. This approach expands model capacity and enhances generalization while maintaining a constant computational cost per token. 
    \item The transition from language-only models to omni-modal systems prevails in newly released models. This transition requires the integration of diverse sub-modules, resulting in increasingly irregular model structures, complicating both parallel programming and optimization. Omni-modal models typically employ a multi-encoder, modal-fusion layer, multi-decoder architecture. This bridging of multiple modalities introduces module-level heterogeneity and dynamic routing loads.
\end{itemize}

\noindent \textbf{Training Paradigms.} The training paradigm changes from traditional pretraining to agentic reinforcement learning (RL). This shift necessitates the co-deployment of diverse tasks, including training, inference, and agent execution. This trend toward heterogenization imposes more stringent requirements on the orchestration of multi-task, multi-model workflows. Concretely, to enhance instruction following, deep reasoning, and complex problem-solving capabilities, state-of-the-art large models increasingly incorporate RL after the initial pre-training phase. Reinforcement learning shifts the workload paradigm from pure training on static datasets to heterogeneous "sample-evaluate-update" tasks involving dynamic data. This workflow typically relies on massive rollout operations and asynchronous actor-learner architectures, where training and inference tasks are executed concurrently. As a result, underlying frameworks must support robust heterogeneous task management, asynchronous task scheduling, and dynamic orchestration capabilities tailored for multi-task, multi-model environments.

\noindent \textbf{Hardware Features.} The Supernode cluster is tailored for the training of LLMs featuring MoE, multimodal, and agentic characteristics. Taking the Huawei Matrix384~\cite{matrix384} (Atlas 900) supernode as an example, we analyze the typical characteristics of supernode clusters:

\begin{itemize}
    \item \textbf{Unified Memory Addressing and Peer-to-Peer Architecture} – Based on the UB interconnect protocol (Lingqu), the system integrates 384 Ascend 910C NPUs and 192 Kunpeng CPUs into a single supernode. Computing, memory, and network resources are decoupled into independent pooled units.
    \item \textbf{Extreme Bandwidth and Ultra-Low Latency} – Traditional architectures rely on PCIe or Ethernet for interconnectivity. The Ascend 384 supernode utilizes a high-efficiency protocol that increases communication bandwidth by 15x compared to traditional server architectures, while reducing single-hop latency from 2$\mu$s to 200ns—a tenfold improvement.
    \item \textbf{Hierarchical and Diverse Network Topologies} – The scale of a single Ascend supernode is projected to increase from 384 to 8,192 cards, with future targets reaching 15,488 cards. To support this massive, scalable interconnect, the supernode employs a hierarchical topology: a 2D full-mesh is created within each rack and extended to another 2D full-mesh across racks, ultimately forming a 4D all-to-all (full-mesh) interconnect topology.
\end{itemize}

\noindent \textbf{Challenges Faced by AI Frameworks.} While supernodes provide superior solutions for rapidly growing model parameters and increasingly complex architectures, realizing their full potential requires addressing several fundamental challenges. To reduce the complexity of algorithm development and tuning while maximizing large-scale training and inference efficiency, AI frameworks face the following critical challenges:

\begin{table}[htbp]
\centering
\begin{minipage}{0.45\textwidth}
\centering
\caption{Strategies by model.}
\begin{tabular}{ll}
\toprule
\textbf{Model \& Algorithm} & \textbf{Strategy} \\
\midrule
\midrule
Dense Transformer & DP, PP, TP, SP \\
\hline
Sparse MoE & DP, PP, TP, SP, EP \\
\hline
Diffusion & DP, FSDP \\
\hline
Long Sequence & SP, CP \\
\hline
RL & MPMD \\
\hline
$\cdots$ & $\cdots$ \\
\bottomrule
\end{tabular}
\label{table:Strategies_by_model}
\end{minipage}
\hfill
\begin{minipage}{0.53\textwidth}
\centering
\caption{Strategies by cluster.}
\begin{tabular}{p{2.6cm}p{3.7cm}}
\toprule
\textbf{Cluster} & \textbf{Strategy} \\
\midrule
\midrule
Single Machine\newline(8 DIE) & TP8,\newline PP for the rest \\
\hline
Single Machine (16 DIE) & High-dimension TP\newline(TP16), reduced PP \\
\hline
8k-Node \newline Hyperplane & Topology-aware TP16, \newline reduced PP \\
\bottomrule
\end{tabular}
\label{table:Strategies_by_cluster}
\end{minipage}
\end{table}

\begin{figure}
    \centering
    \includegraphics[width=\linewidth]{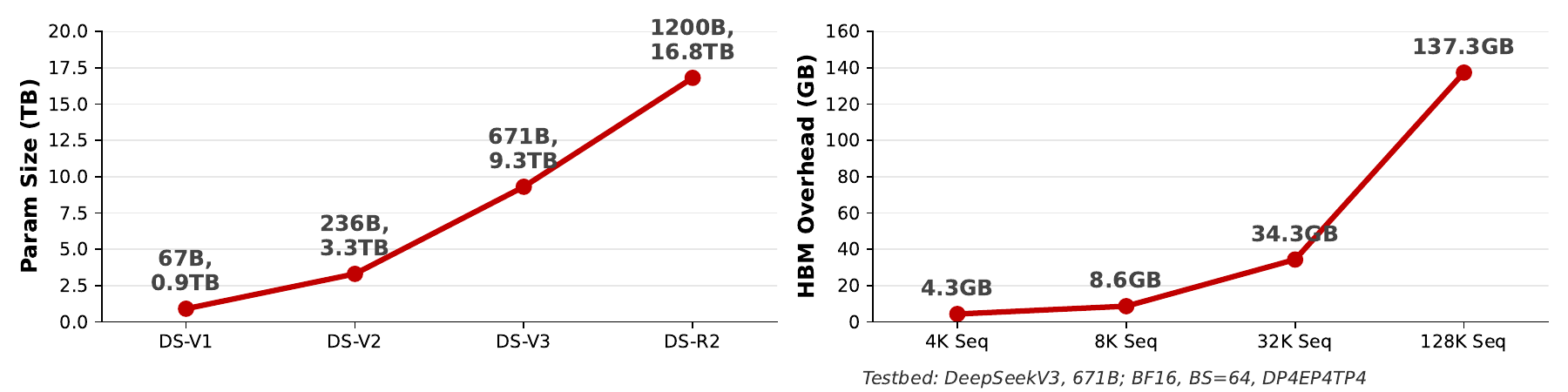}
    \caption{The complexity of storing and managing parameters and intermediate states during model training and inference processes continues to increase.}
    \label{fig:intermediates}
\end{figure}

\begin{itemize}
    \item \textbf{Challenge 1: Abstracting Complex Network Topologies to Simplify Parallel Programming} – As supernodes eliminate inter-node bandwidth bottlenecks, cluster scales can grow to the level of 100,000 cards. Navigating such intricate network topologies requires meticulously designed and optimized parallel strategies. Any change in model architecture or cluster configuration necessitates a complete redesign of these strategies, as shown by the examples in Table~\ref{table:Strategies_by_model} and Table~\ref{table:Strategies_by_cluster}. Empirically, each adaptation and tuning cycle requires 1–2 weeks of effort from senior system optimization engineers and consumes significant cluster resources, leading to operational inefficiency.
    
    \item \textbf{Challenge 2: Implementing Flexible, Fine-Grained Parallelism to Maximize Hardware Utilization} – Current parallel sharding primarily relies on the Single Program, Multiple Data (SPMD) paradigm. However, as models evolve toward MoE and omni-modal architectures, different modules and layers exhibit heterogeneous computational loads. Standard SPMD approaches frequently result in load imbalances; for instance, in omni-modal models with varying sub-module loads, SPMD combined with Pipeline Parallelism (PP) often induces significant pipeline bubbles. Furthermore, MoE models suffer from high communication overhead; in DeepSeek-V3, Expert Parallelism (EP) communication accounts for 17\% of execution time, with a communication masking ratio of only 61\% compared to an ideal target of 90\%.

    \item \textbf{Challenge 3: Managing Intermediate States via Memory Pooling to Optimize Memory Utilization} – As model scales move from hundreds of billions toward tens of trillions of parameters, the complexity of managing intermediate states—including weights, activations, and KV caches—increases exponentially. Supernodes provide a unified DRAM memory pool through memory-semantic interconnects. To maintain performance, intermediate states must be proactively prefetched (swapped) into High Bandwidth Memory (HBM) before use to avoid memory access latency. Manually managing these swapping operations imposes a significant programming and optimization burden on algorithm researchers.
\end{itemize}

\section{Design of a Supernode-Affinity AI Framework}
\label{sec4}
Current optimization efforts in industrial training and inference software are insufficient to fundamentally resolve the systemic challenges imposed by supernodes on AI frameworks. To fully leverage the advantages of supernodes and address the bottlenecks arising from model evolution, AI frameworks must undergo an architectural redesign centered on \textit{supernode affinity}. By integrating the specific architectural characteristics of supernodes, frameworks can provide researchers with a transformative programming experience while significantly enhancing training and inference efficiency. This chapter draws upon the practical implementations of MindSpore to introduce the architectural design and key technologies of a supernode-affinity AI framework.

\subsection{Overall Design}

\begin{figure}
    \centering
    \includegraphics[width=0.9\linewidth]{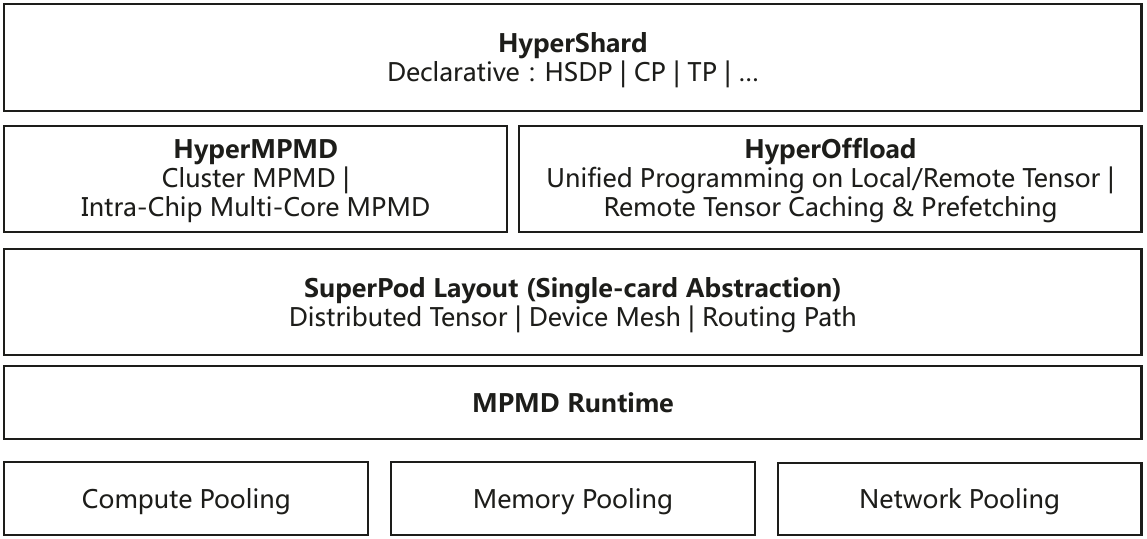}
    \caption{Overall \textit{Hyperparallel} Design.}
    \label{fig:overall}
\end{figure}

The core philosophy of MindSpore's supernode affinity is to abstract the underlying hardware complexity through the AI framework, treating the supernode as a single giant Computer. This approach provides algorithm researchers with a programming and execution experience akin to a single-node system. By embedding complex software engineering tasks—such as parallel sharding and training-inference state management—directly into the framework, MindSpore leverages the peer-to-peer architectural advantages of supernodes to resolve performance bottlenecks in complex scenarios like MoE, omni-modal, and reinforcement learning. To this end, we propose the \textbf{HyperParallel} architecture as shown in Figure~\ref{fig:overall}, which comprises three pivotal components designed to shield users from hardware complexity while maximizing architectural benefits:

\begin{itemize}
    \item \textbf{HyperOffload} – Facilitates the decoupling of computation from state. By utilizing the pooled storage capabilities of the supernode, it resolves the memory bottlenecks introduced by escalating model scales.
    \item \textbf{HyperMPMD} – Transitions parallel sharding from the conventional SPMD paradigm to a fine-grained Multiple Program, Multiple Data (MPMD) approach. This leverages the peer-to-peer computational architecture to provide flexible parallelism for reinforcement learning and omni-modal workloads.
    \item \textbf{HyperShard} – Orchestrates a shift from imperative to declarative programming. It abstracts the intricate topology of the supernode cluster, providing a minimalist, composable programming experience for omni-modal and other advanced models.
\end{itemize}

The operational workflow for model development, training, and inference on supernodes using MindSpore is structured as follows:

\begin{itemize}
    \item \textbf{Step 1: Algorithmic Development} – Researchers develop algorithms through the framework, utilizing HyperShard's declarative programming interface to configure the model's parallel strategies.
    \item \textbf{Step 2: Flexible Parallelism} – HyperMPMD executes flexible, fine-grained parallelism in an MPMD format, optimized specifically for the algorithm and the unique cluster topology.
    \item \textbf{Step 3: Runtime Orchestration} – During execution, HyperOffload performs global orchestration across computation, storage, and networking. It manages the state exchange between High Bandwidth Memory (HBM) and DRAM to fully utilize the supernode's hierarchical memory pool.
\end{itemize}

\subsection{HyperOffload}

\begin{figure}
    \centering
    \includegraphics[width=0.7\linewidth]{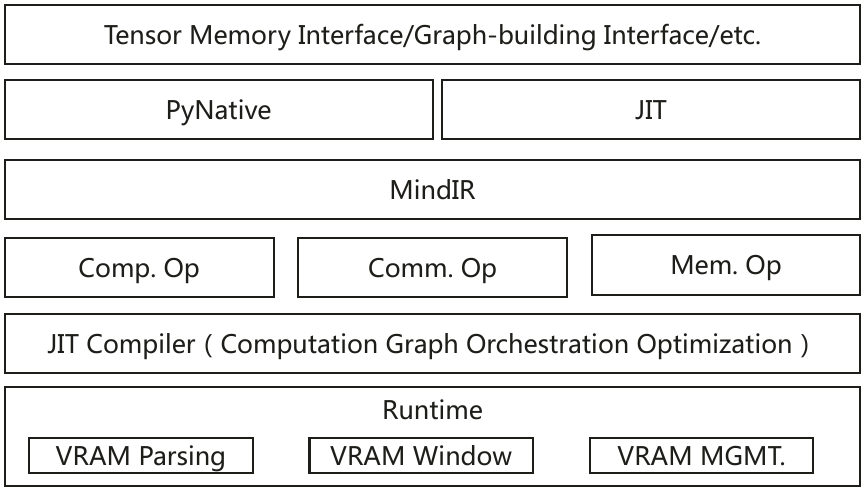}
    \caption{Architecture of \textit{HyperOffload}.}
    \label{fig:HyperOffload}
\end{figure}

The fundamental concept of \textbf{HyperOffload} is to offload model states to an external memory pool, utilizing on-chip memory (HBM) as a high-speed cache. This architecture allows a single accelerator to accommodate significantly larger models. The core challenge lies in automating the state exchange between HBM and DRAM without compromising performance. To address this, HyperOffload employs multi-level cache pipeline scheduling and holistic graph orchestration as shown in Figure~\ref{fig:HyperOffload}.

\begin{itemize}
    \item \textbf{Multi-level Cache Pipeline Scheduling} – HyperOffload utilizes communication hiding techniques to asynchronously prefetch cache blocks required for the next execution phase into the high-speed storage layer before they are requested by computational operators. By integrating model structural characteristics with data access pattern prediction, the system dynamically adjusts prefetch paths to effectively overlap loading latency with computation time.
    \item \textbf{Holistic Graph Orchestration and Scheduling} – HyperOffload abstracts hierarchical cache operations (such as data prefetching and block offloading) into native operators, enabling a unified description of cache read/write and migration behaviors. Leveraging MindSpore’s JIT graph compilation, the framework automatically inserts cache management operators and reorganizes the execution flow after parsing the user network. By performing unified orchestration of cache, computation, and communication operators, the compiler automatically schedules parallel execution chains, eliminating the complexity of manual synchronization point insertion.
\end{itemize}

By fully exploiting the unified large-memory pool of the supernode, HyperOffload significantly relaxes HBM memory constraints during training and inference. This enables the removal of complex $N$D-SPMD parallelism in favor of simple 1D-SPMD Data Parallelism, which reduces the communication overhead associated with state synchronization and further enhances performance.

Empirical results demonstrate substantial improvements:
\begin{itemize}
    \item \textbf{Training Scenarios} – Using the same hardware configuration, the iteration time per step for a Llama-8B model was reduced from 5.2s using traditional methods to 4.08s, representing an approximate 20\% performance increase.
    \item \textbf{Inference Scenarios} – Under identical latency constraints, HyperOffload increased the supported sequence length from 71K to 123K, a 70\% improvement over traditional approaches.
\end{itemize}

\subsection{HyperMPMD}

\begin{figure}
    \centering
    \includegraphics[width=\linewidth]{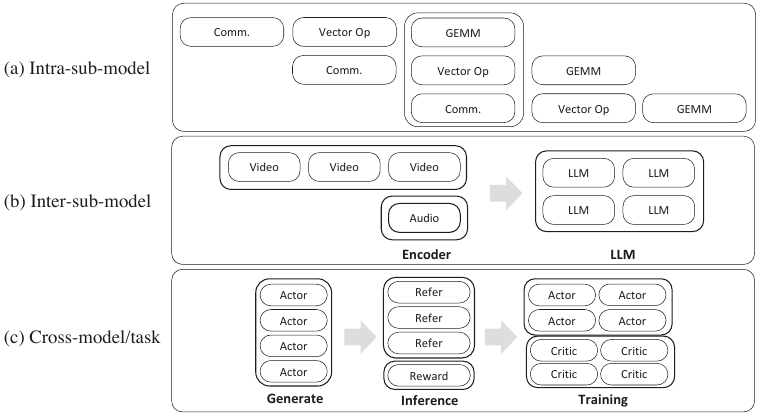}
    \caption{Illustration of HyperMPMD.}
    \label{fig:hypermpmd}
\end{figure}

The core objective of \textbf{HyperMPMD} is to implement a flexible, fine-grained Multiple Program, Multiple Data (MPMD) sharding approach to resolve computational imbalances in scenarios such as omni-modal learning and reinforcement learning, thereby maximizing hardware utilization. HyperMPMD provides MPMD capabilities across three distinct dimensions:

\begin{itemize}
    \item \textbf{Intra-sub-model Core-level Concurrency} -- By sharding model tensors and utilizing intra-card MPMD scheduling for AICube and AIVector tasks, the framework enables fine-grained orchestration of computation-communication overlap (as shown in Figure~\ref{fig:hypermpmd} (a)). This addresses the communication masking challenges inherent in MoE architectures. By synergizing intra-card multi-core parallelism with inter-card parallelism, HyperMPMD increases the communication masking ratio from the traditional 60\% to 90\%.
    \item \textbf{Inter-sub-model Concurrency Balancing} -- The framework decouples subgraphs into independent concurrent tasks (as shown in Figure~\ref{fig:hypermpmd} (b)), utilizing dynamic scheduling to mitigate load imbalances. This effectively eliminates the 10\%--40\% pipeline bubbles typically found in omni-modal or multimodal models caused by heterogeneous sub-module loads, resulting in an overall training performance gain of approximately 15\%.
    \item \textbf{Cross-model Concurrent Scheduling} -- Integrated with the MPMD runtime, the framework provides Single Controller support to perform fine-grained parallel sharding and dynamic scheduling within the supernode's pooled computational resources (as shown in Figure~\ref{fig:hypermpmd} (c)). This enables model-level concurrency and eliminates straggler effects, resolving load imbalances across multi-task reinforcement learning and increasing cluster-wide resource utilization by 15\%.
\end{itemize}

HyperMPMD partitions independent MPMD process groups based on modalities or tasks (e.g., text, image, audio, fusion, and task scheduling groups). Each group executes specialized program logic, communicating via standardized interfaces. The following example illustrates the configuration for an omni-modal MPMD process group. By encapsulating core logic into independent modules and defining \textit{node-to-module} mappings via configuration files as shown in Listing~\ref{code:mpmd_config}, the framework eliminates the need for rigid hard-coding.

\begin{center}
\begin{minipage}{\textwidth} 
\begin{lstlisting}[language=Python, caption={Mappings configuration example.}, label={code:mpmd_config}]
process_groups = {
  "text": {"ranks": [0,1], "module": TextModule, "hardware": "NPU"},
  "image": {"ranks": [2,3], "module": ImageModule, "hardware": "NPU"},
  "audio": {"ranks": [4], "module": AudioModule, "hardware": "CPU"},
  "fusion": {"ranks": [5], "module": CrossModalFusion, "hardware": "NPU"},
  "task_scheduler": {"ranks": [6], "module": TaskPriorityScheduler, 
                     "hardware": "CPU"}
}
\end{lstlisting}
\end{minipage}
\end{center}

\subsection{HyperShard}

\begin{figure}
    \centering
    \includegraphics[width=\linewidth]{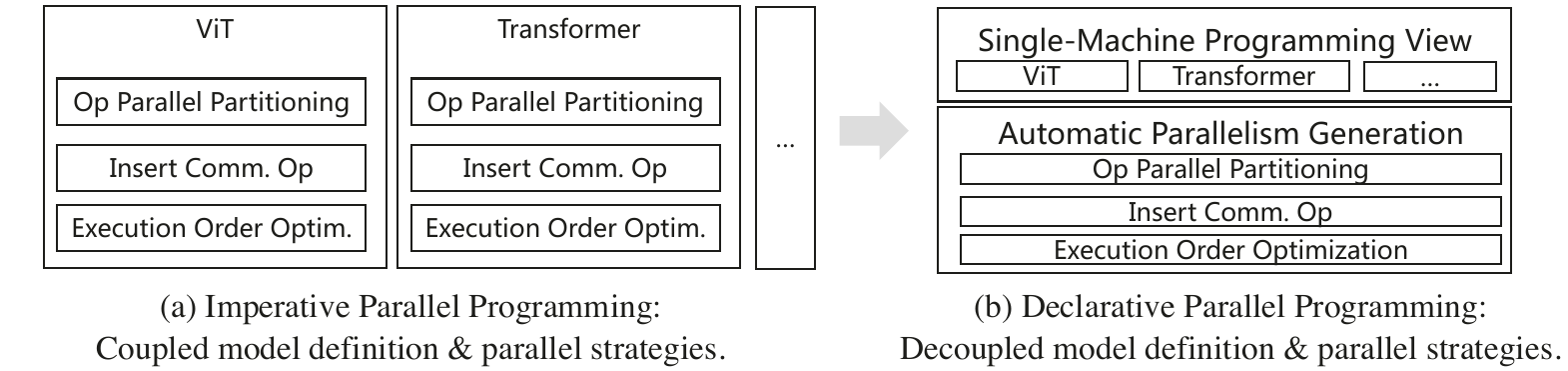}
    \caption{Comparison between declarative and imperative parallel programming.}
    \label{fig:comp_hypershard}
\end{figure}

The fundamental objective of \textbf{HyperShard} is to provide declarative parallel programming capabilities, enabling developers to build models using single-node programming conventions. This approach effectively decouples algorithmic logic from parallel strategies, allowing the latter to be flexibly configured based on network topology. As illustrated in Figure~\ref{fig:comp_hypershard} (a), traditional imperative parallel programming suffers from tight coupling between model definition and parallel strategies, requiring bespoke manual interventions—such as operator sharding, explicit communication operator insertion, and execution sequence optimization—for every model variant. In contrast, Figure~\ref{fig:comp_hypershard} (b) demonstrates the declarative paradigm: researchers develop algorithms from a single-device perspective and merely declare the desired parallel constraints. The framework then automatically generates the underlying parallel strategy, achieving full decoupling.

The primary programming abstraction of \textbf{HyperShard} is the interface \verb|Layout(device_matrix,|   \verb|alias_name, tensor_map)| defined as follows:
\begin{itemize}
    \item \verb|device_matrix| -- Describes the logical arrangement of accelerators within the cluster.
    \item \verb|alias_name| -- Assigns descriptive identifiers to each dimension within the device matrix.
    \item \verb|tensor_map| -- Specifies how each dimension of a given Tensor is partitioned across the device matrix.
\end{itemize}

\begin{center}
\begin{minipage}{0.6\textwidth} 
\begin{lstlisting}[language=Python,label={code:example_hypershard},caption=Hypershard strategy generation example.]
# 4 accelerators as 2*2 device matrix 
device_matrix = (2, 2) 
# alias for each dimension of device matrix
alias_name = ("x", "y")
# assume that tensor.shape=(2, 2)
tensor_map = ("x", "y")
# declare layout
layout = Layout(device_matrix, alias_name)
# generate parallel partitioning strategy
parallel_strategy = layout(tensor_map)
\end{lstlisting}
\end{minipage}
\end{center}

\begin{figure}
    \centering
    \includegraphics[width=1.0\linewidth]{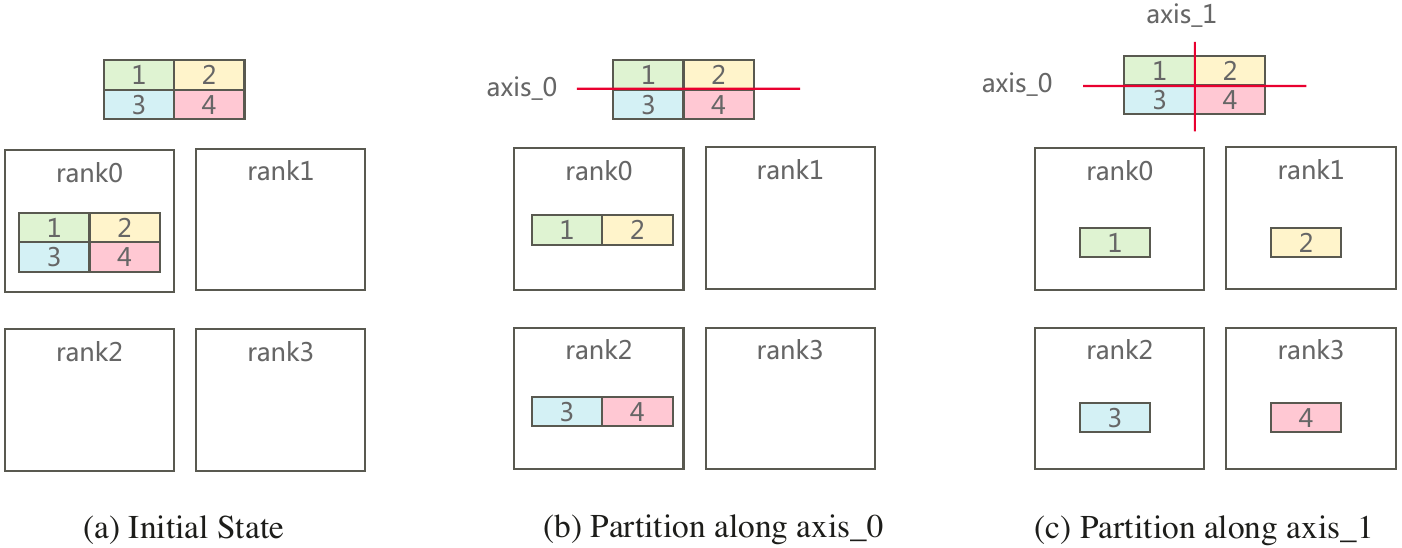}
    \caption{Hypershard partitioning illustration.}
    \label{fig:illus_Hypershard}
\end{figure}

Using the provided code in Listing~\ref{code:example_hypershard} as an example, we illustrate the process of sharding a Tensor across eight accelerators. The automated sharding procedure proceeds in three stages:
\begin{itemize}
    \item \textbf{Figure~\ref{fig:illus_Hypershard} (a): Initialization} -- The framework initially assumes the Tensor resides on a single logical rank (e.g., Rank 0).
    \item \textbf{Figure~\ref{fig:illus_Hypershard} (b): First-Dimension Partitioning} -- Since \verb|tensor_map[0] = "x"|, the framework shards the 0-th dimension of the Tensor along the \verb|"x"| dimension of the \verb|device_matrix|.
    \item \textbf{Figure~\ref{fig:illus_Hypershard} (c): Second-Dimension Partitioning} -- Since \verb|tensor_map[1] = "y"|, the framework shards the 1st dimension of the Tensor along the \verb|"y"| dimension of the \verb|device_matrix|.
\end{itemize}

It is critical to note that this process does not involve physical tensor slicing at compile time. Instead, the framework performs a formal derivation of the parallel strategy (\textit{shard\_strategy}) based on user configuration; the actual execution of the sharding occurs during runtime.

By shielding developers from hardware complexity through the decoupling of algorithms and parallel strategies, \textbf{HyperShard} significantly reduces engineering overhead. The time required for parallelizing new algorithms is reduced to less than one day, while the cycle for parallel strategy optimization is compressed from days to mere hours.

\section{Summary and Discussion}
\label{sec5}
This paper systematically analyzes the evolving trends in model workloads, hardware architectures, and training-inference paradigms, highlighting the challenges that large-scale models and supernode clusters impose on AI frameworks. We identify critical deficiencies in current frameworks regarding parallel sharding, state management, and cluster utilization efficiency. To address these gaps, we propose a design for a supernode-affinity AI framework and implement the \textbf{HyperParallel} architecture based on MindSpore.

Through the integration of three core technologies—\textbf{HyperOffload}, \textbf{HyperMPMD}, and \textbf{HyperShard}—the framework achieves the decoupling of computation from state, enables fine-grained MPMD parallelism, and introduces a declarative programming interface. These innovations yield the following contributions:

\begin{itemize}
    \item \textbf{Abstraction of Complex Topologies} – The framework simplifies the development and tuning pipeline, allowing researchers to develop large-scale models with the same ease of use as single-node programming.
    \item \textbf{Flexible and High-Efficiency Parallelism} – By providing dynamic and granular parallel strategy scheduling, the framework maximizes the utilization of supernode computational resources, effectively resolving load imbalances in MoE, omni-modal, and reinforcement learning scenarios.
    \item \textbf{Unified State Management} – Leveraging unified memory pools to manage intermediate states improves memory utilization, supporting the training and inference of increasingly large models without performance degradation.
\end{itemize}

Empirical results demonstrate that under identical hardware conditions, HyperOffload and HyperMPMD significantly enhance training and inference performance while drastically reducing the time required for algorithmic development and parallel strategy optimization. Looking ahead, the supernode-affinity AI framework can be further extended to support heterogeneous task scheduling, multi-node dynamic scaling, and cross-datacenter collaborative training, providing a robust software infrastructure for the next generation of large-scale model research and industrial deployment.

\section{Acknowledgments}
We would like to extend our sincere gratitude to the following contributors for their invaluable roles in the development of the HyperParallel Framework.

\noindent \textbf{Research \& Engineering}

\noindent Changwei Ma,
Chao Fu,
Chen Li,
Chongming Liu,
Da Lei,
Dawei Fan,
Dongjie Geng,
Fuxin Huang,
Guangpeng Zhang,
Haoran Wang,
Huikang Tang,
Jiahong Qian,
Jiaqi Song,
Jinshan Ding,
Kaisheng Wang,
Mingqi Li,
Nuohang Li,
Qi Guo,
Qian Shu,
Renwei Zhang,
Shanni Li,
Xiangyu Meng,
Xinglei Xu,
Xiong Gao,
Yi Huang,
Yide Wang,
Yifan Yao,
Yixing Feng,
Zhenbang Wang,
Zheng Li,
Zhenzhang Yang,
Zherui Chang,
Zhibo Liang

\bibliographystyle{plain} 
\bibliography{ref}        

\end{document}